\documentclass[a4paper,10pt]{article}
\usepackage[dvips]{graphicx}
\begin{document}
\title{The Study of Shocks in Three-States Driven-Diffusive Systems: A Matrix Product Approach}
\author{F H Jafarpour$^{1}$\footnote{Corresponding author's e-mail:farhad@ipm.ir} \, and S R Masharian$^{2}$ \\ \\
{\small $^1$Bu-Ali Sina University, Physics Department, Hamadan, Iran} \\
{\small $^2$Institute for Advanced Studies in Basic Sciences,
Zanjan, Iran}} \maketitle
\begin{abstract}
We study the shock structures in three-states one-dimensional
driven-diffusive systems with nearest neighbors interactions using a
matrix product formalism. We consider the cases in which the
stationary probability distribution function of the system can be
written in terms of superposition of product shock measures. We show
that only three families of three-states systems have this property.
In each case the shock performs a random walk provided that some
constraints are fulfilled. We calculate the diffusion coefficient
and drift velocity of shock for each family.
\end{abstract}
{\bf PACS \,}{02.50.Ey, 05.20.-y, 05.70.Fh, 05.70.Ln} %
\maketitle %
\section{Introduction}
\label{Sec1}
During recent years, much attention has been paid to the study of
out of equilibrium low-dimensional diffusive particles systems
\cite{sz,sch}. These systems are kept far from equilibrium by
maintaining a steady state particle current. This results in various
critical phenomena including boundary-induced phase transitions,
phase separation and spontaneous symmetry breaking. For instance the
evolution of shocks in these systems is an indication for the phase
separation phenomenon. Macroscopically these shocks are special
solutions of nonlinear hydrodynamic equations for coarse-grained
quantities like the density of particles. Nevertheless in order to
understand the structure of the shocks and also the nonlinear nature
of them it is necessary to study the microscopic dynamics of these
systems in details.\\
A simple two-states lattice gas model, which shows a variety of
interesting critical phenomena including shock formations, is the
Partially Asymmetric Simple Exclusion Process (PASEP) with open
boundaries. In the PASEP the particles are injected into the system
and also extracted from the boundaries while hop in the bulk of a
discrete lattice \cite{dehp}. Recent investigations show that for
the PASEP with open boundaries (and even on an infinite lattice) a
traveling shock with a step-like density profile might evolve in the
system provided that microscopic reaction rates are adjusted
appropriately \cite{fe,ff,dem,bs,ps,kjs}. The shock position in this case
performs a random walk while reflects from the boundaries. It is
also known that for the PASEP multiple shocks can evolve in the
system provided that specific constraints are fulfilled. It turns
out that the PASEP is not the only two-states lattice gas model in
which the dynamics of the single shock under the Hamiltonian of the
system is a random walk. There are also two other systems of this
kind i.e. the Branching-Coalescing Random Walk (BCRW) and also the
Asymmetric Kawasaki-Glauber Process (AKGP). The steady state
probability distribution functions of these three systems are known
to be made of superposition of product shock measures \cite{kjs}.\\
In the context of one-dimensional out of equilibrium
reaction-diffusion systems, the Matrix Product Formalism (MPF) has
shown to be one of powerful technics in order to study of both the
steady state and also the dynamical properties of these systems
\cite{sch}. According to this approach the probability distribution
function of the system at any time (including the steady state) can
be written in terms of the matrix element (for open boundary
conditions) or trace (for periodic boundary conditions) of products
of some non-commuting operators. At large times these operators are
time independent while they can be time dependent at finite times.
For the systems with nearest neighbors interactions these operators
satisfy an algebra which can have finite or infinite-dimensional
representations. For the three above mentioned systems, it has been
shown that the conditions for the existence of a single traveling
shock with random walk behavior in the system are exactly those for
the existence of a two-dimensional representation for their
quadratic algebras \cite{ja1}. On the other hand, it is well known
that the quadratic algebra of the PASEP has also an $n$-dimensional
representation given that some extra constraints on the boundaries and
bulk rates are held \cite{er,ms}. It turns out that these
constraints are exactly those for the existence of $n-1$ consecutive
shocks with random walk dynamics in the system. In this case the
steady state probability distribution function of the PASEP
can be expressed in terms of superposition of $n-1$ shocks.\\
In this paper we consider the most general three-states
reaction-diffusion system with nearest neighbors interactions and
open boundaries. The system is defined on a lattice of length $L$.
Throughout this paper, we assign the letters $A$, $B$ and $E$ to
these states. The $E$'s are associated with the empty sites or holes
in the system while $A$'s and $B$'s are associated with particles of
different types. We define the total density of particles as the sum
of the densities of particles of type $A$ and $B$ at each site. We
also assume that the density of $A$ particles in a given site is
proportional to the density of $B$ particles at the same site. In
terms of the total density of the particles and the density of
vacancies, the three-states system is basically a two-states system.
In present work, we firstly investigate the conditions under which
the stationary probability distribution function of the three-states
system can be expressed in terms of superposition of single shocks.
Our approach will be based on the MPF. From the MPF point of view,
one can simply try to map the resultant algebra of the three-states
system into that of three two-states above mentioned models and find
the constraints on the microscopic reaction rates under which this
mapping is possible. This guarantees that the resultant algebra of
the three-states system has a two-dimensional representation which
in turns indicates that the steady state distribution of the system
can be written in terms of superposition of single shocks. In the
case of mapping the quadratic algebra of the three-states system
into the PASEP's algebra, the constraints are sufficient for
expressing the stationary probability distribution of the
system in terms of superposition of multiple shocks.\\
Secondly we study the dynamics of a single product shock measures in
the three-states systems. The dynamics of a single product shock
measure in two-states systems has already been studied and as we
mentioned in the PASEP, the BCRW and the AKGP the shock performs a
random walk given that some constraints are satisfied. For our
three-states system the conditions under which the quadratic algebra
of the system has a two-dimensional representation are exactly those
for the product shock measure to have a random walk dynamics. We
obtain the diffusion coefficient and the shock drift velocity in each case. \\
This paper is organized as follows: In section \ref{Sec2} we briefly
review the basic concepts of the MPF especially for two and three
states system. The quadratic algebras of the PASEP, the BCRW and
also the AKGP are reviewed in section \ref{Sec3}. We also review the
constraints under which their quadratic algebras have
two-dimensional representations. In section \ref{Sec4} we map the
quadratic algebra of the most general three-states system (in terms
of the total density of the particles and the vacancies) to the
quadratic algebras of these three models and find the constraints
under which the mapping is possible. The dynamics of the shock is
discussed in section \ref{sec5}. The results and conclusion are
brought in the last section.
\section{The Matrix Product Formalism (MPF)}
\label{Sec2}
In this section we review the basic concepts of the MPF introduced
in \cite{dehp}. Let us define $P({\mathcal C};t)$ as the probability
distribution function of any configuration $\mathcal C$ of a
Markovian interacting particle system at the time $t$. The time
evolution of $P({\mathcal C};t)$ can be written as a Schr\"odinger
equation in imaginary time
\begin{equation}
\frac{d}{dt}P({\mathcal C};t)=H P({\mathcal C};t)
\end{equation}
in which $H$ is a stochastic Hamiltonian. The matrix elements of the
Hamiltonian are the transition rates between different
configurations. For the one-dimensional systems defined on a lattice
of length $L$ with nearest neighbors integrations the Hamiltonian
$H$ has the following general form
\begin{equation}
\label{Hamiltonian}
H=\sum_{k=1}^{L-1}h_{k,k+1}+h_1+h_L.
\end{equation}
in which
\begin{equation}
\label{Hamiltoniandet}
\begin{array}{l}
h_{k,k+1}={\cal I}^{\otimes (k-1)}\otimes h \otimes {\cal
I}^{\otimes (L-k-1)} \\ h_1=h^{(l)} \otimes {\cal I}^{\otimes (L-1)}
\\ h_L={\cal I}^{\otimes (L-1)}\otimes h^{(r)}
\end{array}
\end{equation}
For two-states systems ${\cal I}$ is a $2 \times 2$ identity matrix
and $h$ is a $4 \times 4$ matrix for the bulk interactions. The most
general form for the bulk Hamiltonian of a two-states system in the
basis $(00,01,10,11)$ is given by
$$
h = \left( \begin{array}{cccc}
-\omega_{11} &  \omega_{12} &  \omega_{13} &  \omega_{14} \\
 \omega_{21} & -\omega_{22} &  \omega_{23} &  \omega_{24} \\
 \omega_{31} &  \omega_{32} & -\omega_{33} &  \omega_{34} \\
 \omega_{41} &  \omega_{42} &  \omega_{43} & -\omega_{44} \end{array} \right)
$$
and for the boundaries by
$$
h^{(l)} = \left(
\begin{array}{rr}
-\alpha & \gamma \\
\alpha & -\gamma \end{array} \right) \, , \, h^{(r)} =  \left(
\begin{array}{rr}
-\delta & \beta \\
\delta & -\beta \end{array} \right).
$$
In this basis $0$ and $1$ stand for a vacancy and a particle
respectively. Requiring the conservation of probability one should
have $\omega_{ii}=\sum_{j\neq i}\omega_{ji}$. In the most general
form the interaction rates are
\begin{equation}
\label{TWOSrate}
\begin{array}{ll}
\mbox{Diffusion of particles} & \omega_{32}, \; \omega_{23} \\
\mbox{Coalescence of particles} & \omega_{34}, \; \omega_{24} \\
\mbox{Branching of particles} & \omega_{43}, \; \omega_{42} \\
\mbox{Death of particles}  & \omega_{13}, \; \omega_{12} \\
\mbox{Birth of particles} & \omega_{31}, \; \omega_{21}  \\
\mbox{Pair Annihilation and Creation}& \omega_{14}, \; \omega_{41}\\
\mbox{Injection and Extraction of particles at the first site} & \alpha, \; \gamma  \\
\mbox{Injection and Extraction of particles at the last site}  &
\delta, \; \beta.
\end{array}
\end{equation}
In the stationary state we have $H P^*({\cal C})=0$. Let us define
the occupation number $\tau_i$ where $\tau_i=0$ if the site $i$ is
empty and $\tau_i=1$ if it is occupied by a particle. According to
the MPF the stationary probability distribution $P^*(\{\tau_i\})$ is
assumed to be of the form
\begin{equation}
\label{Weigth1} P^*(\{\tau_i\})=\frac{1}{Z_{L}} \langle W \vert
\prod_{i=1}^{L}(\tau_i \textbf{D}+(1-\tau_i)\textbf{E})\vert V
\rangle.
\end{equation}
The function $Z_L$ in (\ref{Weigth1}) is a normalization factor and
can be obtained easily using the normalization condition. The
operators $\textbf{D}$ and $\textbf{E}$ stand for the presence of a
particle and a vacancy at each site. These operators besides the
vectors $\langle W \vert$ and $\vert V \rangle$ should satisfy the
following algebra
\begin{equation}
\label{Twoalg}
\begin{array}{l}
h \left[ \left( \begin{array}{c} \textbf{E} \\
\textbf{D} \end{array} \right) \otimes
\left( \begin{array}{c} \textbf{E} \\
\textbf{D} \end{array} \right) \right]=
\left( \begin{array}{c} \bar{\textbf{E}} \\
 \bar{\textbf{D}} \end{array} \right) \otimes
\left( \begin{array}{c} \textbf{E} \\
 \textbf{D} \end{array} \right) -
\left( \begin{array}{c} \textbf{E} \\
 \textbf{D} \end{array} \right) \otimes
\left( \begin{array}{c} \bar{\textbf{E}} \\
 \bar{\textbf{D}} \end{array} \right), \\
\langle W| \; h^{(l)} \left( \begin{array}{c} \textbf{E} \\
 \textbf{D} \end{array} \right) =
-\langle W| \left( \begin{array}{c} \bar{\textbf{E}} \\
 \bar{\textbf{D}} \end{array} \right),
h^{(r)} \left( \begin{array}{c} \textbf{E} \\
 \textbf{D} \end{array} \right) |V \rangle =
\left(\begin{array}{c} \bar{\textbf{E}} \\
 \bar{\textbf{D}} \end{array} \right) |V \rangle. \nonumber
\end{array}
\end{equation}
The operators $\bar{E}$ and $\bar{D}$ are auxiliary operators and do
not enter in the calculation of physical quantities. \\
For three-states systems the Hamiltonian $H$ is given by
(\ref{Hamiltonian}). In this case ${\mathcal I}$ is a $3 \times 3$
identity matrix, $h$ is a $9 \times 9$ matrix for the bulk
interactions and $h^{(l)}$ ($h^{(r)}$) which stands for the particle
input and output from the left (right) boundary is a $3 \times 3$
matrix. One should note that by requiring the conservation law for
the probabilities the matrix $h$ has only $72$ independent elements.
The boundary matrices $h^{(l)}$ and $h^{(r)}$ have also $6$
independent elements each. Introducing the basis
$(00,01,02,10,11,12,20,21,22)$, in which $0$, $1$ and $2$ stand for
a hole, a particle of type $A$ and a particle of type $B$
respectively, we have
\begin{equation}
\label{Threeham} h=\left( \begin{array}{ccccccccc}
-x_{11} & x_{12} & x_{13} & x_{14} & x_{15} & x_{16} & x_{17} & x_{18} & x_{19}\\
x_{21} & -x_{22} & x_{23} & x_{24} & x_{25} & x_{26} & x_{27} & x_{28} & x_{29}\\
x_{31} & x_{32} & -x_{33} & x_{34} & x_{35} & x_{36} & x_{37} & x_{38} & x_{39}\\
x_{41} & x_{42} & x_{43} & -x_{44} & x_{45} & x_{46} & x_{47} & x_{48} & x_{49}\\
x_{51} & x_{52} & x_{53} & x_{54} & -x_{55} & x_{56} & x_{57} & x_{58} & x_{59}\\
x_{61} & x_{62} & x_{63} & x_{64} & x_{65} & -x_{66} & x_{67} & x_{68} & x_{69}\\
x_{71} & x_{72} & x_{73} & x_{74} & x_{75} & x_{76} & -x_{77} & x_{78} & x_{79}\\
x_{81} & x_{82} & x_{83} & x_{84} & x_{85} & x_{86} & x_{87} & -x_{88} & x_{89}\\
x_{91} & x_{92} & x_{93} & x_{94} & x_{95} & x_{96} & x_{97} & x_{98} & -x_{99}\\
\end{array} \right)
\end{equation}
for the bulk Hamiltonian and also
\begin{equation}
\label{Threehambl} h^{(l)} = \left(
\begin{array}{cccc}
-(\alpha_1+\alpha_2) & \gamma_1 & \gamma_2  \\
\alpha_1 & -(\gamma_1+\alpha_3) & \gamma_3  \\
\alpha_2 & \alpha_3 & -(\gamma_2+\gamma_3)
\end{array} \right)
\end{equation}
and
\begin{equation}
\label{Threehambr} h^{(r)} = \left( \begin{array}{cccc}
-(\delta_1+\delta_2) &\beta_1 & \beta_2  \\
\delta_1 & -(\beta_1+\delta_3) & \beta_3  \\
\delta_2 & \delta_3 & -(\beta_2+\beta_3)
\end{array} \right)
\end{equation}
for the boundaries. As we mentioned the conservation of
probabilities in (\ref{Threeham}) requires $x_{ii}=\sum_{j\neq
i}x_{ji}$ for $i,j=1,\cdots,9$. At the left boundary the particles
of kind $A$ and $B$ are injected into the system with the rates
$\alpha_1$ and $\alpha_2$ respectively. These particles can also
leave the system from there with the rates $\gamma_1$ and
$\gamma_2$. There are also possibilities for changing the particle
types at this boundary. The particles of type $A$ ($B$) are
converted to the particles of type $B$ ($A$) with rate $\alpha_3$
($\gamma_3$). The same processes can also take place at the right
boundary for the particles of type $A$ and $B$ i.e. the particle
injection (with the rates $\delta_1$ and $\delta_2$), the particle
extraction (with the rates $\beta_1$ and $\beta_2$) and also the
particle type conversion (with the rates $\delta_3$ and $\beta_3$).
In the three-states case let us define some new notation. We
introduce two occupation numbers, $\tau_i$ and $\theta_i$, for each
site of the lattice, where $\tau_i=1$ if the site $i$ is occupied by
a particle of type $A$ and $0$ otherwise. Similarly, $\theta_i=1$ if
the site $i$ is occupied by a particle of type $B$ and $0$
otherwise. As the particles are assumed to be subjected to an
excluded-volume interaction, only one of $\tau_i$ and $\theta_i$ can
be nonzero and the configuration of the system $\mathcal C$ is
uniquely defined by the set of occupation numbers $\{
\tau_i,\theta_i \}$. According to the MPF the stationary probability
distribution $P^*(\{ \tau_i, \theta_i \})$ for an open system is
assumed to be of the form
\begin{equation}
\label{Weigth2} P^*(\{ \tau_i, \theta_i \})=\frac{1}{Z_{L}} \langle
W \vert \prod_{i=1}^{L}(\tau_i \textbf{A}+\theta_i \textbf{B} +
(1-\tau_i-\theta_i) \textbf{E})\vert V \rangle.
\end{equation}
The function $Z_L$ in (\ref{Weigth2}) is again a normalization
factor and can be obtained easily using the normalization condition.
The operators \textbf{E}, \textbf{A} and \textbf{B} stand for the
presence of a vacancy, a particle of kind $A$ and a particle of kind
$B$ at each site. These operators besides the vectors $\langle W
\vert$ and $\vert V \rangle$ should satisfy the following algebra
\begin{equation}
\label{Threealg}
\begin{array}{l}
\label{MPA} h \left[ \left( \begin{array}{c} \textbf{E} \\
\textbf{A} \\ \textbf{B}
\end{array} \right) \otimes
\left( \begin{array}{c} \textbf{E} \\
\textbf{A} \\ \textbf{B} \end{array} \right) \right]=
\left( \begin{array}{c} \bar{\textbf{E}} \\
 \bar{\textbf{A}} \\  \bar{\textbf{B}} \end{array} \right) \otimes
\left( \begin{array}{c} \textbf{E} \\
\textbf{A} \\ \textbf{B} \end{array} \right) -
\left( \begin{array}{c} \textbf{E} \\
\textbf{A} \\ \textbf{B} \end{array} \right) \otimes
\left( \begin{array}{c} \bar{\textbf{E}} \\
 \bar{\textbf{A}} \\  \bar{\textbf{B}} \end{array} \right), \\
\langle W \vert \; h^{(l)} \left( \begin{array}{c} \textbf{E} \\
\textbf{A} \\ \textbf{B} \end{array} \right) =
-\langle W \vert \left( \begin{array}{c} \bar{\textbf{E}} \\
 \bar{\textbf{A}} \\  \bar{\textbf{B}} \end{array} \right),
h^{(r)} \left( \begin{array}{c} \textbf{E} \\
\textbf{A} \\ \textbf{B} \end{array} \right) \vert V \rangle =
\left(\begin{array}{c} \bar{\textbf{E}} \\
 \bar{\textbf{A}} \\  \bar{\textbf{B}} \end{array} \right) \vert V \rangle. \nonumber
\end{array}
\end{equation}
As before the operators $\bar{\textbf{E}}$, $\bar{\textbf{A}}$ and
$\bar{\textbf{B}}$ are auxiliary operators and do not enter in the
calculation of physical quantities. Using the Hamiltonian of the
system given by (\ref{Threeham})-(\ref{Threehambr}) and (\ref{MPA})
the quadratic algebra associated with the most general three-states
reaction-diffusion model can be obtained. In order to calculate the
mean values of the physical quantities, such as the mean density of
particles at each site, one should either find a matrix
representation for the algebra or use the commutation relation of
the algebra directly.\\
In the next section we will review the two-states systems in which
the Hamiltonian of the system develop a single product shock measure
as a random walk provided that the reaction rates fulfill some
constraints.
\section{Shocks in two-states systems}
\label{Sec3}
In \cite{kjs} exact traveling wave solutions are obtained for three
families of one-dimensional two-species reaction-diffusion models
with open boundaries. It has been shown that for these models the
stationary probability distribution functions can be written as a
linear combination of Bernoulli shock measures defined on a lattice
of length $L$ as
\begin{equation}
\label{BSM}
\vert k \rangle= \left( \begin{array}{c} 1-\rho_1 \\
\rho_1 \end{array} \right)^{\otimes k} \otimes \left( \begin{array}{c} 1-\rho_2 \\
\rho_2 \end{array} \right)^{\otimes L-k} \; \;,\; \; 0\leq k \leq L
\end{equation}
provided that some constraints on the reaction rates are fulfilled.
In (\ref{BSM}) $\rho_1$ and $\rho_2$ are the densities of particles
at the left and the right domains of the shock position $k$
respectively. The time evolution of (\ref{BSM}) generated by
the Hamiltonian of the above-mentioned models is given by
\begin{equation}
\label{shock} \frac{d}{dt} \vert k \rangle = d_l \vert k-1 \rangle +
d_r \vert k+1 \rangle - (d_l+d_r) \vert k \rangle \; \; , \; \; 0 <
k < L
\end{equation}
which is a simple single-particle random walk equation for the
position of the shock $k$. The shock positions hops to the left and
to the right with the rates $d_l$ and $d_r$ respectively. In the
following we briefly review the PASEP, the BCRW and the AKGP and the
conditions under which we have an exact traveling shock solution for
the model \cite{kjs}. These models are defined on an integer lattice
of length $L$, each site of the lattice $i$ ($1 \leq i \leq L$) is
either empty or occupied by at most one particle. By applying the
standard MPF and defining the operators $\textbf{D}$ and
$\textbf{E}$ as the operators associated with the existence of
particles and vacancies, one finds the following results for the
above mentioned models:\\
For the PASEP the non-vanishing rates in
(\ref{TWOSrate}) are $\omega_{32}$,
$\omega_{23}$, $\alpha$, $\beta$, $\gamma$ and $\delta$. Using
(\ref{Twoalg}) the quadratic algebra of the PASEP has now the
following quadratic form
\begin{equation}
\label{PASEPalg}
\begin{array}{l}
\omega_{23}\textbf{DE}-\omega_{32}\textbf{ED}=(\omega_{23}-\omega_{32})(\textbf{D+E})\\
\langle W \vert (\alpha \textbf{E} - \gamma \textbf{D} ) = \langle W \vert\\
(\beta \textbf{D}- \delta \textbf{E})\vert V \rangle=\vert V
\rangle.
\end{array}
\end{equation}
The auxiliary operators $\bar{\textbf{E}}$ and $\bar{\textbf{D}}$
are real numbers proportional to unity in this case. It is known
that the Hamiltonian of the PASEP has an eigenvector with zero
eigenvalue which can be written in terms of superposition of single
product shock measures. The non-vanishing rates together with these
densities should satisfy the following conditions
\begin{equation}
\label{PASEPcond}
\begin{array}{l}
\frac{\rho_2(1-\rho_1)}{\rho_1(1-\rho_2)}=\frac{\omega_{23}}{\omega_{32}}\\
\alpha(1-\rho_1)-\gamma\rho_1=\rho_1(1-\rho_1)(\omega_{23}-\omega_{32}) \\
\beta
\rho_2-\delta(1-\rho_2)=\rho_2(1-\rho_2)(\omega_{23}-\omega_{32}).
\end{array}
\end{equation}
In the bulk of the lattice the shock position hops to the left with the rate
$d_l=(\omega_{23}-\omega_{32})\frac{\rho_1(1-\rho_1)}{\rho_2-\rho_1}$
and to the right with the rate
$d_r=(\omega_{23}-\omega_{32})\frac{\rho_2(1-\rho_2)}{\rho_2-\rho_1}$.
The shock velocity and the shock diffusion coefficient will then can
be read as $v_s=d_r-d_l$ and $D_s=(d_r+d_l)/2$. It has also been
shown that under the conditions (\ref{PASEPcond}) the algebra (\ref{PASEPalg})
has a two-dimensional representation \cite{ja1}.
The $n$-dimensional representation of (\ref{PASEPalg}) describes the stationary
distribution of the PASEP with $n-1$ consecutive shocks. It has been
shown that an $n$-dimensional irreducible representation for any
finite $n$ exists provided that the following constraint is
satisfied by the bulk and the boundary rates
\begin{equation}
\label{MultiCond}
(\frac{\omega_{32}}{\omega_{23}})^{1-n}=\kappa_{+}(\alpha,\gamma)\kappa_{+}(\beta,\delta)
\end{equation}
in which we have defined
\begin{equation}
\label{kappa}
\kappa_{+}(u,v)=\frac{-u+v+1+\sqrt{(u-v-1)^2+4uv}}{2u}.
\end{equation}
For the BCRW the non-vanishing parameters in (\ref{TWOSrate}) are
$\omega_{24}$, $\omega_{42}$, $\omega_{34}$, $\omega_{43}$,
$\omega_{32}$, $\omega_{23}$, $\alpha$, $\beta$ and $\gamma$. The
dynamics of a single product shock measure under the Hamiltonian of
the BCRW is a random walk provided that
\begin{equation}
\label{BCRWcond}
\begin{array}{l}
\frac{1-\rho_1}{\rho_1}=\frac{\omega_{24}+\omega_{34}}{\omega_{42}+\omega_{43}}\\
\frac{1-\rho_1}{\rho_1}=\frac{\omega_{23}}{\omega_{43}}\\
\gamma= \frac{1-\rho_1}{\rho_1}\alpha + (1-\rho_1) \omega_{32}-
\frac{1-\rho_1}{\rho_1} \omega_{43} + \rho_1 \omega_{34}
\end{array}
\end{equation}
while the density of the particles at the right hand side of the
shock position is zero $\rho_2=0$. The shock position hops to the
left and right with the hopping rates
$d_l=(1-\rho_1)\omega_{32}+\rho_1 \omega_{34}$ and
$d_r=\omega_{43}/\rho_1$. The shock drift velocity and diffusion
coefficient are then $v_s=d_r-d_l$ and $D_s=(d_r+d_l)/2$. Using the
standard MPF the quadratic algebra of the BCRW is obtained to be
\begin{equation}
\label{BCRWalg}
\begin{array}{l}
\bar{\textbf{E}} \textbf{E}- \textbf{E}\bar{\textbf{E}}=0\\
\omega_{23}\textbf{DE}+\omega_{24}\textbf{D}^2-(\omega_{32}+\omega_{42})
\textbf{ED}=\bar{\textbf{E}}\textbf{D}-\textbf{E}\bar{\textbf{D}}\\
-(\omega_{23}+\omega_{43})\textbf{DE}+\omega_{34}\textbf{D}^2+\omega_{32}
\textbf{ED}=\bar{\textbf{D}}\textbf{E}-\textbf{D}\bar{\textbf{E}}\\
\omega_{43}\textbf{DE}-(\omega_{24}+\omega_{34})\textbf{D}^2+\omega_{42}
\textbf{ED}=\bar{\textbf{D}}\textbf{D}-\textbf{D}\bar{\textbf{D}}\\
\langle W \vert (\alpha \textbf{E}-\gamma \textbf{D})=\langle W
\vert\bar{\textbf{E}}=-\langle W \vert\bar{\textbf{D}}\\
\beta \textbf{D} \vert V \rangle =\bar {\textbf{E}} \vert V \rangle
=-\bar {\textbf{D}} \vert V \rangle.
\end{array}
\end{equation}
It has been shown that the steady state probability distribution
function of the BCRW defined by (\ref{Weigth1}), can be described by
a two-dimensional representation of (\ref{BCRWalg}) given that
the conditions (\ref{BCRWcond}) are fulfilled \cite{ja1}.\\
For the AKGP, the non-vanishing parameters in (\ref{TWOSrate}) are
$\omega_{12}$, $\omega_{13}$, $\omega_{42}$, $\omega_{43}$,
$\omega_{32}$, $\alpha$ and $\beta$. The particles are allowed only
to enter the system from the first site with the rate $\alpha$ and
leave it from the last site of the lattice with the rate $\beta$. In
this case we have $\rho_1=1$ and $\rho_2=0$. There are no additional
constraints on the rates for this model. The dynamics of shock
measure generated by the Hamiltonian of the system will be a simple
random walk on the lattice and the shock position hopping rates are
$d_l=\omega_{13}$ and $d_r=\omega_{43}$. The shock drift velocity
and diffusion coefficient can now be easily calculated. It has also
been shown that the quadratic algebra of the AKGP given by
\begin{equation}
\label{AKGPalg}
\begin{array}{l}
\omega_{13}\textbf{DE}+\omega_{12}\textbf{ED}= \bar{\textbf{E}} \textbf{E}-\textbf{E} \bar{\textbf{E}}\\
-(\omega_{12}+\omega_{32}+\omega_{42}) \textbf{ED}=\bar{\textbf{E}}\textbf{D}-\textbf{E}\bar{\textbf{D}}\\
-(\omega_{13}+\omega_{43})\textbf{DE}+\omega_{32}\textbf{ED}=\bar{\textbf{D}}\textbf{E}-\textbf{D}\bar{\textbf{E}}\\
\omega_{43}\textbf{DE}+\omega_{42}\textbf{ED}=\bar{\textbf{D}}\textbf{D}-\textbf{D}\bar{\textbf{D}}\\
\langle W \vert \alpha \textbf{E}=\langle W \vert\bar{\textbf{E}}=-\langle W \vert\bar{\textbf{D}}\\
\beta \textbf{D} \vert V \rangle =\bar {\textbf{E}} \vert V \rangle
=-\bar {\textbf{D}} \vert V \rangle
\end{array}
\end{equation}
has a two-dimensional representation. The stationary state of the
system can be written in terms of superposition of product shock measures (\ref{BSM}).\\
In the following section, by defining the total density of
particles, we map the quadratic algebra of our three-states system
into (\ref{PASEPalg}), (\ref{BCRWalg}) and (\ref{AKGPalg}). We find
the conditions under which such mapping is possible.

\section{Mapping of algebras}
\label{Sec4}
In what follows we consider the most general three-states system
consists of two species of particles and vacancies and look for the
constraints under which the steady state probability distribution
function of the system can be written in terms of superposition of
product shock measures. It turns out that the dynamics of such
product shock measure under the Hamiltonian of the system is a
random walk under the same constraints. In the most general case in
a three-states system, the particles belong to two different types
$A$ and $B$ and the vacancies which will be denoted by $E$. We
define the total density of particles as the sum of the densities of
$A$ and $B$ particles at each lattice site. We assume that at any
given site of the lattice the mean occupation value of the particles
of type $A$ is always proportional to the mean occupation value of
the particles of type $B$. From the MPF point of view, this means
that the operator $\textbf{A}$, associated with the presence of an
$A$ particle at a given site, should be proportional to the the
operator $\textbf{B}$ which is associated with the presence of a $B$
particle i.e. $\textbf{A}=\frac{1}{r}\textbf{B}$ in which $r$ is a
real and positive parameter. This implies that the operators
$\textbf{A}$ and $\textbf{B}$ commute with each other but not
necessarily with $\textbf{E}$ which in turn means the stationary
probability for occurrence of any configurations of type $\cdots
EEEAAABABABBBABEE\cdots$ does not depend on the exact position of
$A$'s and $B$'s in each block surrounded by $E$'s rather to the
number of them. We should emphasis that such constraint does not
result in a trivial two-states model even at the microscopic level
(see \cite{ts1} for instance). Now by defining the total density of
particles operator $\textbf{D}$ as
$\textbf{D}:=\textbf{A}+\textbf{B}$ the associated quadratic algebra
of the three-states system (\ref{Threealg}) can be written in terms
of the two operators $\textbf{D}$ and $\textbf{E}$ and the auxiliary
operators $\bar{\textbf{E}}$, $\bar{\textbf{A}}$ and
$\bar{\textbf{B}}$. We also assume that
$\bar{\textbf{A}}=\frac{1}{r}\bar{\textbf{B}}$ and define
$\bar{\textbf{D}}:=\bar{\textbf{A}}+\bar{\textbf{B}}$. Under these
assumptions one finds a quadratic algebra in term of the four
operators $\textbf{D}$, $\bar{\textbf{D}}$, $\textbf{E}$ and
$\bar{\textbf{E}}$. This algebra can be regarded as an associated
quadratic algebra of a two-states system with open boundaries. Now
the question is \textit{"what do we get if we try to map this
quadratic algebra into the quadratic algebras associated with the
PASEP, the BCRW and that of the AKGP?"}. Such mapping will indeed
impose some constraints on the reaction rates of the three-states
system. The first result is that the quadratic algebra of the
three-states system can have two-dimensional representations by
imposing extra constraints. This means that the stationary
distribution of the three-states system, when it is described in
terms of the total density of particles and the density of
vacancies, can be written in terms of superposition of single
product shock measures. Moreover, if the total density of the
particles has a step-function structure, it will evolve similar to a
random walker in continues time under the Hamiltonian of the system.
In the following, we map the algebra of the three-states system in
terms of $\textbf{D}$ and $\textbf{E}$ into (\ref{PASEPalg}),
(\ref{BCRWalg}) and (\ref{AKGPalg}) and investigate the outcomes. As
we mentioned, this will impose some conditions on the reaction rates
of our three-states system i.e. $x_{ij}$'s introduced in
(\ref{Threeham}) and also the boundary rates in (\ref{Threehambl})
and (\ref{Threehambr}). The constraints (\ref{PASEPcond}) and
(\ref{BCRWcond}) in the case of mapping into the PASEP's and the
BCRW's algebra will then be applied to the total density of
particles and the newly defined boundary rates. These constraints
will guarantee that the dynamics of (\ref{BSM}) will be a simple
random walk and that the steady state of the system can be written
in terms of superposition of these shocks.
\subsection{Mapping into the PASEP's algebra}       
Here we introduce the conditions under which the quadratic algebra
of the three-states system can be mapped into (\ref{PASEPalg}). One
should note that the quadratic algebra of the PASEP does not contain
any quadratic terms of types $\textbf{E}^2$ or $\textbf{D}^2$. Since
we have defined $\textbf{A}=\frac{1}{1+r}\textbf{D}$ and
$\textbf{B}=\frac{r}{1+r}\textbf{D}$; therefore, the quadratic
algebra of the three-states system should not contain the quadratic
terms of types $\textbf{A}^2$, $\textbf{B}^2$ or $\textbf{E}^2$.
This imposes some constraints on the matrix elements of
(\ref{Threeham}). On the other hand, in order to have a quadratic
term of type $\textbf{DE}$ one should only have the quadratic terms
of types $\textbf{AE}$ and $\textbf{BE}$. All other combinations of
$\textbf{A}$, $\textbf{B}$ and $\textbf{E}$ should also be
eliminated in the quadratic algebra of the three-sates system. By
applying these constraints we have found that the most general
Hamiltonian of the three-states system (\ref{Threeham}) should have
the following form in order to be mapped into (\ref{PASEPalg})
\begin{equation}
\label{ThreehamPASEP} h=\left( \begin{array}{ccccccccc}
0 & 0 & 0 & 0 & 0 & 0 & 0 & 0 & 0\\
0 & -x_{22} & x_{23} & x_{24} & 0 & 0 & x_{27} & 0 & 0\\
0 & x_{32} & -x_{33} & x_{34} & 0 & 0 & x_{37} & 0 & 0\\
0 & x_{42} & x_{43} & -x_{44} & 0 & 0 & x_{47} & 0 & 0\\
0 & 0 & 0 & 0 & -x_{55} & x_{56} & 0 & x_{58} & x_{59}\\
0 & 0 & 0 & 0 & x_{65} & -x_{66} & 0 & x_{68} & x_{69}\\
0 & x_{72} & x_{73} & x_{74} & 0 & 0 & -x_{77} & 0 & 0\\
0 & 0 & 0 & 0 & x_{85} & x_{86} & 0 & -x_{88} & x_{89}\\
0 & 0 & 0 & 0 & x_{95} & x_{96} & 0 & x_{98} & -x_{99}\\
\end{array} \right)
\end{equation}
in which $x_{ii}=\sum_{j\neq i}x_{ji}$ for $i=2,\cdots,9$. All of
the boundary rates in (\ref{Threehambl}) and (\ref{Threehambr}) are
non-zero in this case. We assume that the auxiliary operators
$\bar{\textbf{D}}$ and $\bar{\textbf{E}}$ are proportional to real
numbers $\bar{\textbf{d}}$ and $\bar{\textbf{e}}$. In this case one
finds that $\bar{\textbf{d}}=-\bar{\textbf{e}}$. By defining
$\bar{\textbf{e}}:=\omega_{32}-\omega_{23}$ one can see that the
algebra of the three-states system can be mapped into
(\ref{PASEPalg}) provided that
\begin{equation}
\label{mapPASEP1}
\begin{array}{l}
r x_{68}+r^2 x_{69}+x_{65}=r x_{66}\\
r x_{86}+r^2 x_{89}+x_{85}=r x_{88}\\
r(x_{96}+x_{98})+x_{95}=r^2 x_{99}.
\end{array}
\end{equation}
One should also define the bulk parameters as follows
\begin{equation}
\label{mapPASEP2}
\begin{array}{ll}
\omega_{23} & := x_{24}+r x_{27} = x_{37}+\frac{1}{r} x_{34} \\
& =x_{44}-r x_{47} = x_{77}-\frac{1}{r}x_{74}\\
\omega_{32} & := x_{42}+r x_{43} = x_{73}+\frac{1}{r} x_{72}\\
& =x_{22}-r x_{23} = x_{33}-\frac{1}{r} x_{32}
\end{array}
\end{equation}
and the boundary rates as
\begin{equation}
\label{mapPASEP3}
\begin{array}{l}
\alpha:=\frac{(1+r)\alpha_1}{\omega_{23}-\omega_{32}}=\frac{\alpha_2(1+r)}{r(\omega_{23}-\omega_{32})}\\
\beta:=\frac{\beta_1+\delta_3-r\beta_3}{\omega_{23}-\omega_{32}}=\frac{r(\beta_2+\beta_3)-\delta_3}{r(\omega_{23}-\omega_{32})}\\
\gamma:=\frac{r(\gamma_2+\gamma_3)-\alpha_3}{r(\omega_{23}-\omega_{32})}=\frac{\gamma_1+\alpha_3-r\gamma_3}{\omega_{23}-\omega_{32}}\\
\delta:=\frac{(1+r)\delta_1}{\omega_{23}-\omega_{32}}=\frac{(1+r)\delta_2}{r(\omega_{23}-\omega_{32})}.
\end{array}
\end{equation}
The conditions in (\ref{PASEPcond}) should now be applied to the
total density of the particles and the new boundary parameters
defined in (\ref{mapPASEP3}).\\
In \cite{ts1} the authors have introduced a three-states system with
a Hamiltonian similar to (\ref{ThreehamPASEP}) and shown that the
dynamics of a single product shock measure under this Hamiltonian
will be a random walk provided that the constraints similar to
(\ref{mapPASEP1})-(\ref{mapPASEP3}) are fulfilled. It should be
mentioned that since (\ref{PASEPalg}) has also finite-dimensional
representations, multiple shock structures may evolve in this system
under the condition (\ref{MultiCond}). Mapping the quadratic algebra
of the three-states system into the PASEP's algebra not only
confirms the results obtained in \cite{ts1} but also prove the
possibility of evolving multiple shocks in the same system which has
not been studied before. One should also note that the models
studied in \cite{rs} and \cite{ja2} are special cases of the system
studied here.
\subsection{Mapping into the BCRW's algebra}                      %
In order to map the quadratic algebra of the three-states system
into (\ref{BCRWalg}) we define the total density operator
$\textbf{D}$ and rewrite the algebra in terms of the two operators
$\textbf{D}$ and $\textbf{E}$ as before. Nevertheless, the auxiliary
operators are not assumed to be real numbers is this case. One
should note that the quadratic algebra of the BCRW does not contain
any quadratic term of type $\textbf{E}^2$. This means that all of
the entries in the first column of (\ref{Threealg}) should be equal
to zero. All other combinations of $\textbf{A}$, $\textbf{B}$ and
$\textbf{E}$ are allowed; therefore, the most general Hamiltonian
should have the following form
\begin{equation}
\label{ThreehamBCRW}
h=\left( \begin{array}{ccccccccc}
0 & 0 & 0 & 0 & 0 & 0 & 0 & 0 & 0\\
0 & -x_{22} & x_{23} & x_{24} & x_{25} & x_{26} & x_{27} & x_{28} & x_{29}\\
0 & x_{32} & -x_{33} & x_{34} & x_{35} & x_{36} & x_{37} & x_{38} & x_{39}\\
0 & x_{42} & x_{43} & -x_{44} & x_{45} & x_{46} & x_{47} & x_{48} & x_{49}\\
0 & x_{52} & x_{53} & x_{54} & -x_{55} & x_{56} & x_{57} & x_{58} & x_{59}\\
0 & x_{62} & x_{63} & x_{64} & x_{65} & -x_{66} & x_{67} & x_{68} & x_{69}\\
0 & x_{72} & x_{73} & x_{74} & x_{75} & x_{76} & -x_{77} & x_{78} & x_{79}\\
0 & x_{82} & x_{83} & x_{84} & x_{85} & x_{86} & x_{87} & -x_{88} & x_{89}\\
0 & x_{92} & x_{93} & x_{94} & x_{95} & x_{96} & x_{97} & x_{98} & -x_{99}\\
\end{array} \right)
\end{equation}
in which, for the conservation of the probability, $x_{ii}$'s are
defined so that the sum of the elements of each column is zero. The
reason that the entries of the first row in (\ref{ThreehamBCRW}) are
taken to be equal to zero comes from the fact that it gives a linear
combination of operators with positive coefficients equal to zero.
This cannot be true unless the coefficients (reaction rates on the
first row) are zero. We now can map the quadratic algebra of the
three-states system into (\ref{BCRWalg}) provided that the nonzero
parameters of the BCRW's algebra are defined as follows
\begin{equation}
\label{mapBCRW1}
\begin{array}{lll}
\omega_{23}& :=x_{24}+rx_{27}=\frac{1}{r}(x_{34}+rx_{37})&\\
\omega_{24}& :=\frac{1}{1+r}(x_{25}+r(x_{26}+x_{28})+r^2 x_{29}) &\\
      & =\frac{1}{r(1+r)}(x_{35}+r(x_{36}+x_{38})+r^2 x_{39})&\\
\omega_{32}& :=x_{42}+rx_{43}=\frac{1}{r}(x_{72}+r x_{73}) & \\
\omega_{34}& :=\frac{1}{1+r}(x_{45}+r(x_{46}+x_{48})+r^2x_{49}) & \\
      & =\frac{1}{r(1+r)}(x_{75}+r(x_{76}+x_{78})+r^2 x_{79})&\\
\omega_{42}& :=(1+r)(x_{52}+rx_{53})=\frac{1+r}{r}(x_{62}+rx_{63})& \\
      & =\frac{1+r}{r}(x_{82}+rx_{83})=\frac{1+r}{r^2}(x_{92}+rx_{93})& \\
\omega_{43}& :=(1+r)(x_{54}+rx_{57})=\frac{1+r}{r}(x_{64}+rx_{67}) & \\
      & =\frac{1+r}{r}(x_{84}+rx_{87})=\frac{1+r}{r^2}(x_{94}+rx_{97}). &
\end{array}
\end{equation}
There are also some extra constraints that should be fulfilled
\begin{equation}
\label{mapBCRW2}
\begin{array}{ll}
x_{22}-rx_{23} & = \frac{1}{r}(rx_{33}-x_{32})\\
x_{44}-rx_{47} & = \frac{1}{r}(rx_{77}-x_{74})\\
x_{55}-r(x_{56}+x_{58})-r^2x_{59}& =\frac{1}{r}(-x_{65}+r(x_{66}-x_{68})-r^2x_{69})\\
& =\frac{1}{r}(-x_{85}+r(x_{88}-x_{86})-r^2x_{89})\\
&=\frac{1}{r^2}(-x_{95}-r(x_{96}+x_{98})+r^2x_{99}).
\end{array}
\end{equation}
The boundary rates in this case should be defined as follows
\begin{equation}
\label{mapBCRW3}
\begin{array}{l}
\alpha:=\alpha_1(1+r)=\alpha_1+\alpha_2\\
\beta:=\beta_1-r\beta_3+\delta_3=\frac{1}{1+r}(\beta_1+r\beta_2)\\
\gamma:=\gamma_1+\alpha_3-r\gamma_3=\frac{1}{1+r}(\gamma_1+r\gamma_2)\\
\delta:=\delta_1=\delta_2=0.
\end{array}
\end{equation}
The last constraint in (\ref{mapBCRW3}) indicates that there is no
particle injection from the right boundary. The conditions given by
(\ref{BCRWcond}) should now be applied to the total density of
particles and newly defined parameters in (\ref{mapBCRW1}) and
(\ref{mapBCRW3}). These conditions and definitions are enough for
writing the steady state probability distribution function of the
model in terms of two-dimensional representation of the algebra.
\subsection{Mapping into the AKGP's algebra}               %
In the case of mapping the quadratic algebra of our three-states
system into the AKGP's algebra, the Hamiltonian of the three-states
system (\ref{Threeham}) should have the following form
\begin{equation}
\label{ThreehamAKGP}
h=\left( \begin{array}{ccccccccc}
0 & x_{12} & x_{13} & x_{14} & 0 & 0 & x_{17} & 0 & 0\\
0 & -x_{22} & x_{23} & 0 & 0 & 0 & 0 & 0 & 0\\
0 & x_{32} & -x_{33} & 0 & 0 & 0 & 0 & 0 & 0\\
0 & x_{42} & x_{43} & -x_{44} & 0 & 0 & x_{47} & 0 & 0\\
0 & x_{52} & x_{53} & x_{54} & 0 & 0 & x_{57} & 0 & 0\\
0 & x_{62} & x_{63} & x_{64} & 0 & 0 & x_{67} & 0 & 0\\
0 & x_{72} & x_{73} & x_{74} & 0 & 0 & -x_{77} & 0& 0\\
0 & x_{82} & x_{83} & x_{84} & 0 & 0 & x_{87} & 0 & 0\\
0 & x_{92} & x_{93} & x_{94} & 0 & 0 & x_{97} & 0 & 0\\
\end{array} \right).
\end{equation}
Since (\ref{AKGPalg}) does not contain any quadratic terms of types
$\textbf{D}^2$ and $\textbf{E}^2$, the quadratic algebra of the
three-states system should not contain the terms $\textbf{A}^2$,
$\textbf{B}^2$, $\textbf{AB}$, $\textbf{BA}$ and $\textbf{E}^2$;
therefore, the first, the fifth, the sixth and the last two columns
of (\ref{ThreehamAKGP}) are chosen to be zero; however, all other
combinations of $\textbf{A}$, $\textbf{B}$ and $\textbf{E}$ are
allowed in this case. In order to map the quadratic algebra of the
three-states model in terms of the operators $\textbf{D}$,
$\textbf{E}$, $\bar{\textbf{D}}$ and $\bar{\textbf{E}}$ into the
AKGP's algebra given by (\ref{AKGPalg}) the nonzero parameters in
(\ref{ThreehamAKGP}) should be related to the nonzero parameters of
the AKGP as follows
\begin{equation}
\begin{array}{lll}
\omega_{12} & := & \frac{1}{1+r}(x_{12}+r x_{13})\\
\omega_{13} & := & \frac{1}{1+r}(x_{14}+r x_{17})\\
\omega_{32} & := & x_{42}+r x_{43}\\
       & = &  \frac{1}{r}(x_{72}+r x_{73})\\
\omega_{42} & := & (1+r)(x_{52}+r x_{53})\\
       & =  & \frac{1+r}{r}(x_{62}+r x_{63})\\
       & =  & \frac{1+r}{r}(x_{82}+r x_{83})\\
       & =  & \frac{1+r}{r^2}(x_{92}+r x_{93})\\
\omega_{43} & := & (1+r)(x_{54}+r x_{57})\\
       & =  & \frac{1+r}{r}(x_{64}+r x_{67})\\
       & =  & \frac{1+r}{r}(x_{84}+r x_{87})\\
       & =  & \frac{1+r}{r^2}(x_{94}+r x_{97})
\end{array}
\end{equation}
besides the following constraints
\begin{equation}
\label{mapAKGP1}
\begin{array}{l}
x_{22}-r x_{23} = x_{33}- \frac{x_{32}}{r}\\
x_{44}-r x_{47} = x_{77} -\frac{x_{74}}{r}
\end{array}
\end{equation}
in which $x_{ii}$'s for $i=2,3,4$ and $7$ are defined according to
the conservation of probabilities. For the boundary rates one should
have $\alpha_3=r \gamma_3$. On the other hand we should define
\begin{equation}
\label{mapAKGP2}
\begin{array}{l}
\alpha:=(1+r)\alpha_1=\alpha_1+\alpha_2\\
\beta:=\beta_1-r\beta_3+\delta_3=\frac{1}{1+r}(\beta_1+r\beta_2)\\
\gamma:=\gamma_1=\gamma_2=0\\
\delta:=\delta_1=\delta_2=0.
\end{array}
\end{equation}
These conditions are enough in order to write the steady state of
the three-states system in terms of two-dimensional matrices which satisfy
(\ref{AKGPalg}).
\section{Dynamics of shock}                                       
\label{sec5}                                                      
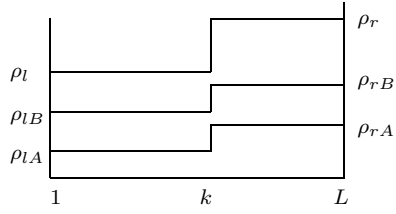
\begin{figure}
\begin{center}
\begin{picture}(120,70)
\put(10,10){\line(0,1){60}} \put(10,10){\line(1,0){110}}
\put(10,20){\line(1,0){60}} \put(10,35){\line(1,0){60}}
\put(70,35){\line(0,1){10}} \put(70,45){\line(1,0){50}}
\put(70,20){\line(0,1){10}} \put(70,30){\line(1,0){50}}
\put(120,10){\line(0,1){66}} \put(10,50){\line(1,0){60}}
\put(70,50){\line(0,1){20}} \put(70,70){\line(1,0){50}}
\put(10,0){\footnotesize 1} \put(66,0){\footnotesize $k$}
\put(116,0){\footnotesize $L$} \put(125,27){\footnotesize
$\rho_{rA}$} \put(125,45){\footnotesize $\rho_{rB}$}
\put(125,68){\footnotesize $\rho_{r}$} \put(-5,17){\footnotesize
$\rho_{lA}$} \put(-5,32){\footnotesize $\rho_{lB}$}
\put(-5,48){\footnotesize $\rho_{l}$}
\end{picture}
\caption[fig]{Sketch of a single shock at sites $k$ on a lattice of
length $L$. The density profile of $A$ particles is always assumed
to be proportional to the density profile of $B$ particles both on
the left hand and the right hand sides of the shock position.}
\label{fig}
\end{center}
\end{figure}

Let us represent a product shock measure with the shock position on
the site $k$ for our three-states system as follows
\begin{equation}
\label{BSM-three} \vert k \rangle = \left( \begin{array}{c}
                    1-\rho_{lA}-\rho_{lB} \\
                    \rho_{lA} \\
                    \rho_{lB}
                  \end{array} \right )^{\otimes k}\otimes\left( \begin{array}{c}
                    1-\rho_{rA}-\rho_{rB} \\
                    \rho_{rA} \\
                    \rho_{rB}
                  \end{array} \right )^{\otimes L-k}
\end{equation}
in which $\rho_{lA}$ ($\rho_{rA}$) and $\rho_{lB}$ ($\rho_{rB}$) are
the densities of $A$ and $B$ particles on the left (right) hand
side of the shock position respectively. According to our assumption
that the density of $A$ particles is always proportional to that of
$B$ particles, one should have $\rho_{lB}=r\rho_{lA}$ and
$\rho_{rB}=r\rho_{rA}$. Now defining the total density of particles
on the sides of the shock position as $\rho_{l}$ and $\rho_{r}$, we
have $\rho_l=(1+r)\rho_{lA}$ and $\rho_r=(1+r)\rho_{rA}$. Sketch of
such shock measure can be seen in figure \ref{fig}.
In the first case (mapping the quadratic algebra of the three-states
system into that of the PASEP) the shock (\ref{BSM-three}) has a
random walk dynamics under (\ref{ThreehamPASEP}) provided that the
constraints (\ref{mapPASEP1})-(\ref{mapPASEP3}) are satisfied. In this case we obtain
\begin{eqnarray}
& & \rho_l = \frac{1}{1+\kappa_{+}(\alpha,\gamma)}  \\
& & \rho_r =
\frac{\kappa_{+}(\beta,\delta)}{1+\kappa_{+}(\beta,\delta)}
\end{eqnarray}
in which we have used (\ref{kappa}) and the boundary rates $\alpha$,
$\beta$, $\gamma$ and $\delta$ are defined according to
(\ref{mapPASEP3}). These boundary rates besides the total density of
particles satisfy the following constraint
\begin{equation}
\begin{array}{l}
\frac{\rho_r(1-\rho_l)}{\rho_l(1-\rho_r)}=\frac{\omega_{23}}{\omega_{32}}\\
\alpha(1-\rho_l)-\gamma\rho_l=\rho_l(1-\rho_l)(\omega_{23}-\omega_{32}) \\
\beta
\rho_r-\delta(1-\rho_r)=\rho_r(1-\rho_2)(\omega_{23}-\omega_{32}).
\end{array}
\end{equation}
The shock position hops to the left and the right with the rates
$\frac{1-\rho_l}{1-\rho_r}w_{32}$ and $\frac{\rho_r}{\rho_l}w_{32}$
respectively.\\
In the second case (mapping the quadratic algebra of the
three-states system to that of the BCRW) the shock (\ref{BSM-three}) performs a random
walk again under (\ref{ThreehamBCRW}) provided that
(\ref{mapBCRW1})-(\ref{mapBCRW3})
are fulfilled. The total density of the particle on the right hand
side of the shock is zero. The shock position in this case hops to
the left and to the right with the rates
$(1-\rho_l)\omega_{32}+\rho_l \omega_{34}$ and $\omega_{43}/\rho_l$
respectively in which $\omega_{32}$, $\omega_{34}$ and $\omega_{43}$
are defined in (\ref{mapBCRW1}). The total density of the particles
on the left hand side of the shock $\rho_l$, the boundary parameters
$\alpha$ and $\delta$, besides the non-zero parameters defined in
(\ref{mapBCRW1}) should now satisfy the following constraints
\begin{equation}
\begin{array}{l}
\frac{1-\rho_l}{\rho_l}=\frac{\omega_{24}+\omega_{34}}{\omega_{42}+\omega_{43}}\\
\frac{1-\rho_l}{\rho_l}=\frac{\omega_{23}}{\omega_{43}}\\
\gamma= \frac{1-\rho_l}{\rho_1}\alpha + (1-\rho_l) \omega_{32}-
\frac{1-\rho_l}{\rho_l} \omega_{43} + \rho_l \omega_{34}.
\end{array}
\end{equation}
In the third case of mapping the quadratic algebra of the
three-states system to the algebra of the AKGP the shock
(\ref{BSM-three}) performs a random walk under the Hamiltonian
(\ref{ThreehamAKGP}). In this case the total density of the
particles of the left hand side of the shock position is equal to
unity while it is zero on the right hand side of the shock position.
The shock position hops to its leftmost (rightmost) site with the
rate $w_{13}$($w_{43}$). There is no need to introduce any extra
constraints more than those in (\ref{mapAKGP1}) in this case.
\section{Concluding remarks}                                      
\label{sec6}                                                      
In this paper we have considered the most general Hamiltonian for
the Markovian three-states systems (two species of particles besides
the vacancies) with open boundaries. From the MPF point of view, the
quadratic algebra of this system is generated by three operators
$\textbf{A}$, $\textbf{B}$, and $\textbf{E}$ associated with the
existence of two species of particles and the vacancies. We have
then assumed that the density of particles are proportional to each
other. By defining the total density of the particles as sum of the
densities of the particles, this algebra transforms into a quadratic
algebra in terms of the total density operator $\textbf{D}$ and
$\textbf{E}$. Now this algebra can be treated as the quadratic
algebra of a two-states system. We have shown that under certain
conditions this algebra can be mapped to the quadratic algebras of
the PASEP, the BCRW and that of the AKGP. Such mapping has two immediate
results: The stationary state of our three-states system can be
written in terms of superposition of product shock measures and that
the dynamics of a single product shock measure given by the
Hamiltonian of the system is simply a random walk.\\
One can simply check that the conditions which are necessary for
mapping the algebra of the three-states system into the PASEP's
algebra, are exactly those introduced in \cite{ts1} but obtained
from different approach. In \cite{ts1} the authors have introduced a
single product shock measure and found the conditions under which it
has a simple random walk dynamics under the Hamiltonian
(\ref{PASEPalg}). In this paper we have shown that (\ref{PASEPalg})
is not the only way one can define the Hamiltonian in order to have
such property. There are actually two other ways which have been
introduced and studied here. In the same reference the authors have
only studied the dynamics of a single shock. We have shown that, at
least for the case of mapping the algebra of the three-states system
into the PASEP's algebra, multiple shocks can evolve in the
system.\\
One should note that the procedure introduced in this paper is not
the only way one can write the quadratic algebra of a three-states
system in terms of the two operators $\textbf{D}$ and $\textbf{E}$.
It is worth studying the case where the operators $\textbf{A}$ and
$\textbf{B}$ are not proportional but related to each other. Our
approach can also be generalized to the systems with more than three
states at each lattice site. This is under our investigations
\cite{jm}.

\end{document}